\documentclass[12pt]{article}
\usepackage[utf8]{inputenc}
\usepackage[T1]{fontenc}

\usepackage{setspace} 
\usepackage[vmargin = 1.5in, hmargin = 1.5in]{geometry}

\usepackage[usenames,dvipsnames,svgnames]{xcolor}
\usepackage{amsmath, amssymb, amsfonts, graphicx, tikz,pdflscape, mathtools, amsthm, upgreek, bm,pgfplots,csvsimple,multirow, multicol, booktabs}
\usepackage{verbatim}
\usepackage{natbib}
\usepackage{accents}

\usepackage{needspace}
\def\citeapos#1{\citeauthor{#1}'s (\citeyear{#1})}
\usepackage{comment}
\usepackage[inline]{enumitem}
\usepackage{tocvsec2}
\usepackage{threeparttable}
\usetikzlibrary{calc, cd}
\usepackage[skip = 10pt]{caption}
\allowdisplaybreaks[1]
\allowdisplaybreaks[1]

\definecolor{Blue}{RGB}{86,180,233}
\definecolor{Orange}{RGB}{230,159,0}
\definecolor{Green}{RGB}{0,158,115}
\definecolor{GmailBlue}{RGB}{42, 93, 176} 
\usepackage[
pagebackref,
colorlinks=true,
citecolor= GmailBlue,
linkcolor=GmailBlue,
urlcolor = GmailBlue
]{hyperref}

\newcommand{\bibtexorder}[1]{}

\usepackage{pgfplots}
\pgfplotsset{compat=newest}
\usepgfplotslibrary{groupplots}
\usepackage{tikz}
\usetikzlibrary{matrix,calc,shapes,arrows.meta,positioning}
\pgfplotsset{width = \textwidth/2}
\tikzstyle{hollow}=[circle,draw,inner sep=1.5]
\tikzstyle{solid}=[circle,draw,inner sep=1.5,fill=black]
\pgfplotsset{compat = newest}

\usepackage[capitalize,noabbrev]{cleveref}

\newtheoremstyle{breakital}
{}
{}
{\itshape}
{}
{\bfseries}
{}
{\newline}
{}

\theoremstyle{breakital}
\newtheorem{thm}{Theorem}
\newtheorem*{theorem*}{Theorem}
\newtheorem*{cor*}{Corollary}

\newtheorem{prop}{Proposition}

\crefname{prop}{Proposition}{Propositions}
\crefname{thm}{Theorem}{Theorems}
\crefname{lem}{Lemma}{Lemmas}

\newtheoremstyle{break}
{}
{}
{}
{}
{\bfseries}
{}
{\newline}
{}

\theoremstyle{break}

\crefname{as}{Assumption}{Assumptions}

\theoremstyle{definition}

\newtheorem*{rem*}{Remark}

\numberwithin{lem2}{section}
\crefname{lem2}{Lemma}{Lemmas}

\def\l{\lambda}

\def\R{\mathbf{R}}

\def\CC{\mathcal{C}}

\def\EE{\mathcal{E}}
\def\FF{\mathcal{F}}

\DeclareMathOperator*{\argmax}{argmax}

\DeclareMathOperator{\ext}{ext}

\title{Bauer's Maximum Principle for Quasiconvex Functions}
\author{Ian Ball}
\date{8 May 2023}

\begin{document}
	
\maketitle

Bauer's maximum principle is increasingly applied to optimization problems in economic theory; see, e.g., \cite{KleinerEtal2021} and  \cite{ArielieEtal2023}. This note shows that in Bauer's maximum principle, the assumed convexity of the objective function can be relaxed to quasiconvexity. 

\begin{thm}[Quasiconvex maximum principle]  \label{res:quasiconvex}
	Let $K$ be a nonempty compact, convex subset of a Hausdorff locally convex space. 
	Any upper semicontinuous, quasiconvex function from $K$ to $[-\infty, \infty)$  achieves its maximum at some extreme point of $K$. 
\end{thm}

The maximum principle with a convex objective is attributed to \cite{Bauer1958}. I first sketch the standard proof of Bauer's maximum principle with a convex objective function; see \citet[p.~298]{AliprantisBorder2006} and  \citet[p.~658]{Ok2007}. Recall that in any vector space, a subset $A$ of a set $B$
is \emph{$B$-extremal} if the following holds: for any 
distinct points $x, y \in B$, if the open interval $(x,y)$ intersects $A$, then $x$ and $y$ are both in $A$.\footnote{For $x \neq y$, the interval  $(x,y)$ is defined to equal the set $\{ \l x + (1 - \l) y : 0 < \l < 1 \}$.} 

The standard proof of Bauer's maximum principle proceeds as follows.   Let $f  \colon K \to [-\infty, \infty)$  be upper semicontinuous and convex. Let $E$ denote the set of maximizers of $f$ over $K$. 

\begin{enumerate}
	\item \label{it:extremal} Check that  the set $E$ is nonempty, closed, and $K$-extremal.
	\item \label{it:Zorn} Use Zorn's Lemma to show that there exists a minimal nonempty, closed, $K$-extremal subset of $E$. Call this set $E_0$. 
	\item \label{it:singleton} Show that $E_0$ is a singleton by arguing as follows. If $E_0$ contains at least two points, then by the the Hahn--Banach theorem there exists a continuous linear functional that is not constant on $E_0$.  The set of maximizers of this linear functional over $E_0$ is a nonempty, closed proper subset of $E_0$ that  can be shown to be $K$-extremal, contrary to the minimality of $E_0$. 
\end{enumerate}

The convexity of  $f$ is used only in part \ref{it:extremal} to show that  the set $E$ of maximizers is $K$-extremal. \citet[p.~300]{AliprantisBorder2006}  show that the same proof goes through if $f$ is assumed to be explicitly quasiconvex, rather than convex.\footnote{A function $f \colon K \to [-\infty, \infty)$ is  \emph{explicitly quasiconvex} if  $f( \l  x + (1 - \l) y) \leq  \max\{ f(x), f(y)\}$ for all distinct $x, y \in K$ and $\l \in (0,1)$, with strict inequality if $f(x) \neq f(y)$. Explicit quasiconvexity is an ordinal property; convexity is not.} Explicit quasiconvexity is  strictly weaker than convexity but strictly stronger than quasiconvexity.  


If $f$ is only quasiconvex, then the set $E$ of maximizers is not necessarily $K$-extremal. But $E$ is \emph{$K$-semi-extremal}, i.e., for any  distinct points $x, y \in K$, if the open interval $(x,y)$ intersects $E$, then at least one of $x$ and $y$ is in $E$. Equivalently, $K \setminus E$ is convex. The proof of  \cref{res:quasiconvex}  below follows the same three-step structure, with  ``$K$-semi-extremal'' in place of ``$K$-extremal'', but step \ref{it:singleton} requires a new argument. The proof is inspired by the proof of the Krein--Milman theorem in \cite{Conway}, which is attribued to \cite{leger1968convexes}. \cref{res:quasiconvex} seems to be known in the mathematical literature, but it is difficult to find a reference with a complete proof.\footnote{\citet[p.~75]{Holmes1975} states \cref{res:quasiconvex}, but does not provide a proof. \citet[p.~425]{Stenger2021} claim that \cref{res:quasiconvex} can be proven by following the proof of Bauer's maximum principle for convex functions in \citet[p.~102]{ChoquetEtal1969}. But this is not correct; claim (iv) in the proof \citet[p.~103]{ChoquetEtal1969} does not hold if the objective is assumed only to be quasiconvex. 
	
The key step in the proof of \cref{res:quasiconvex} is proving that a closed $K$-semi-extremal subset of $K$ contains an extreme point of $K$. This claim is proven in \cite{AndersonPoulson1968} and \cite{Pryce1969}. Those papers do not mention Bauer's maximum principle, and the proofs are quite different than the proof given here.}

\begin{proof}[Proof of Theorem~\ref{res:quasiconvex}] Let $f \colon K \to [-\infty, \infty)$ be upper semicontinuous and quasiconvex.  Since $K$ is compact and $f$ is upper semicontinuous,  the function $f$ achieves its maximum over $K$. Denote the maximum value by $M$, and let $E$ denote the set of maximizers. Since $f$ is upper semicontinuous, the set $E$ is closed. We claim that $E$ is $K$-semi-extremal. Let $x$ and $y$ be distinct points in $K$ such that the open interval $(x,y)$ intersects $E$. Choose $z$ in $(x,y) \cap E$. We have 
	\[
		M = f(z) \leq \max \{ f(x), f(y)\} \leq M,
	\]
	where the middle inequality uses quasiconvexity. It follows that $\max \{ f(x), f(y) \} = M$, so at least one of $x$ and $y$ is in $E$.

	
Let $\EE$ be the collection of all nonempty, closed, $K$-extremal subsets of $E$. Partially order $\EE$ by set inclusion. We show that $\EE$ has a minimal element. Clearly, $\EE$ is nonempty since it contains $E$. Let $\CC$ be a nonempty chain in $\EE$. We claim that the intersection $C = \bigcap \CC$ is a member of $\EE$.  Every member of $\EE$ is  compact and nonempty, so $C$ is compact and nonempty. We claim that $C$ is $K$-semi-extremal.  Let $x$ and $y$ be distinct points in $K$ such that the open interval $(x,y)$ intersects $C$. For each $E'$ in $\CC$, the open interval $(x,y)$ must also intersect $E'$ (since $C \subseteq E'$), so at least one of $x$ and $y$ is in $E'$ (since $E'$ is $K$-semi-extremal). Since $\CC$ is a chain, it follows that at least one of $x$ and $y$, say $x$, is a member of  every set in $\EE$.\footnote{Suppose not. Then there exists some $E_1 \in \CC$ that does not contain $x$  and some $E_2 \in \CC$ that does not contain $y$. Since $\CC$ is a chain, either $E_1 \subseteq E_2$ or $E_2 \subseteq E_1$. The smaller of the two sets  $E_1$ and $E_2$ contains neither $x$ nor $y$, which is a contradiction.} Thus, $x$ is in $C$. We conclude that $C$ is in $\EE$.  We have shown that every chain in $\EE$ has a lower bound in $\EE$. By Zorn's Lemma, $\EE$ has a minimal element, which we denote $E_0$.

To complete the proof, we show that $E_0$ contains a single point, which is therefore an extreme point of $K$. Suppose for a contradiction that $E_0$ contains at least two points.  We show that there is a set  $E_0' \in \EE$ that is strictly smaller than $E_0$, contrary to the minimality of $E_0$. There are two cases. 
\begin{enumerate} [label = \roman*.]
	\item Suppose that $E_0$ is $K$-extremal. 
Since the space is Hausdorff locally convex,  there exists a convex, relatively open subset $V$ of $K$ that contains some point of $E_0$ and excludes another point of $E_0$. Let $E_0' = E_0 \setminus V$. By construction, $E_0'$ is a nonempty, closed proper subset of $E_0$.  We claim that $E_0'$ is $K$-semi-extremal. Let $x$ and $y$ be distinct points of $K$ such that the open interval $(x,y)$ intersects $E_0'$ (and hence $E_0$). Since $E_0$ is $K$-extremal, both $x$ and $y$ are in $E_0$. We claim that at least one of $x$ and $y$ is in $E_0'$. Otherwise $x$ and $y$ are both in $V$.  Since $V$ is convex, the interval  $(x,y)$ is a subset of $V$ and hence cannot intersect $E_0$, a contradiction.

\item Suppose that $E_0$ is not $K$-extremal. Then there exist points $x$ in $K \setminus E_0$ and $y$ in $E_0$  and a scalar $\l$ in $(0,1)$ such that $\l x +(1 - \l) y$ is in $E_0$. Let $E_0'$ be the set of all points $z$ in $K$ for which $\l x + (1 - \l ) z $ is in $E_0$.  In particular, $E_0'$ contains $y$. Since $K \setminus E_0$ is relatively open and convex, it follows that $K \setminus E_0'$ is relatively open and convex, i.e., $E_0'$ is closed and $K$-semi-extremal. To see that $E_0'$ is a proper subset of $E_0$, consider the sequence $(z_n)$ defined by $z_1 = y$ and $z_{n+1} =  \l x + (1 - \l)  z_{n}$ for all $n \geq 1$. This sequence converges to $x$ in $K \setminus E_0$, so only finitely many  terms are in $E_0$. The last term in $E_0$ is in  $E_0 \setminus E_0'$.   \qedhere
\end{enumerate}

\end{proof}

To conclude, we comment on the assumption that $K$ is convex. The original statement of the maximum principle in \cite{Bauer1958} does not assume that the domain $K$ is convex. In fact, the standard proof of Bauer's principle with a convex objective, sketched above, does not use the assumption that  $K$ is convex. That proof therefore establishes a stronger maximum principle in which $K$ is not assumed to be convex (and the objective is explicitly quasiconvex, as discussed above).\footnote{A function defined on a nonconvex domain  is convex/(explicitly) quasiconvex if the defining inequalities hold for 
for convex combinations that are in the domain.}

By contrast, the convexity of $K$ in \cref{res:quasiconvex} cannot be dropped. \cite{Haberl2004} gives a simple example of a nonempty compact, nonconvex subset $K$ of $\R^2$ and a closed $K$-semi-extremal subset $E$ of $K$ that does not contain an extreme point of $K$. The indicator function for $E$ is upper semicontinuous and quasiconvex, but it does not achieve its maximum at any extreme point of $K$.\footnote{ \cite{Bereanu1974} claims to prove a generalized maximum principle for quasiconvex functions on domains that are not necessarily convex. The example in \cite{Haberl2004} is a counterexample to that result. \cite{Haberl2004} discusses the error in \citeapos{Bereanu1974} proof.}

\bibliographystyle{ecta}
\bibliography{Bauer_lit.bib}

\end{document}